\documentstyle[aps,pre,multicol,epsf]{revtex}

\def\D{\partial}
\def\grad{\nabla}
\def\div{\nabla \cdot}

\def\eq{\begin{eqnarray}}
\def\qe{\end{eqnarray}}

\def\eqnn{\begin{eqnarray*}}
\def\qenn{\end{eqnarray*}}

\def\bem{\bm{m}}
\def\bn{\bm{n}}
\def\bq{\bm{q}}
\def\br{\bm{r}}
\def\bu{\bm{u}}

\def\simge{\;\lower3pt\hbox{$\stackrel{\textstyle >}{\sim}$}\;}
\def\simle{\;\lower3pt\hbox{$\stackrel{\textstyle <}{\sim}$}\;}

\def\bm#1{\mbox{\boldmath $#1$}}
\def\tensor#1{{\sf #1}}

\def\lrS#1{\left(#1\right)}
\def\lrangle#1{\left\langle #1 \right\rangle}

\def\f#1#2{\frac{#1}{#2}}
\def\der#1#2{\f{\D #1}{\D #2}}
\def\fder#1#2{\f{\delta #1}{\delta #2}}

\newcommand{\tW}{\tensor{W}}

\title  {Elastic Effects in Disordered Nematic Networks}
\author {Nariya Uchida}
\address{Department of Physics, Kyoto University, Kyoto 606, Japan}
\date   {\today}
\begin{document}
\draft

\bibliographystyle{prsty}

\maketitle
\begin{abstract}
Elastic effects in a model of disordered nematic elastomers 
are numerically investigated in two dimensions.
Networks crosslinked in the isotropic phase
exhibit unusual soft mechanical response against stretching.
It arises from gradual alignment of orientationally 
correlated regions that are elongated along the director.
A sharp crossover to a macroscopically 
aligned state is obtained on further stretching. 
The effect of random internal stress is also discussed.
\end{abstract}


\begin{multicols}{2} 

Nematic elastomers and gels exhibit rich mechanical effects 
due to elasticity-orientation coupling \cite{deGennesCRAS,Warner}.
While a considerable number of theoretical studies
has been directed to homogeneous systems,
nematic elastomers are often in a highly non-uniform 
polydomain state, in which the correlation length for 
the director orientation is typically of micron scales. 
Polydomain networks show unusual non-linear elastic response against 
stretching,
often with an extremely low stress over a large interval of 
strain~\cite{Finkelmann,Finkelmann94,Bergmann-etal,Zubarev-etal,Clarke2,Clarke-Terentjev}.
As the strain is increased, the directors gradually rotate 
toward the direction of stretching until a macroscopically aligned state 
is attained. This structural change 
is called the polydomain-monodomain (P-M) transition. 
Attempting to describe the presumably equilibrium polydomain textures,
Terentjev and coworkers~\cite{Clarke1,Fridrikh-Terentjev,Yu-etal}
proposed a random-field model analogous to those for random 
anisotropy magnets. They argued that crosslinkers of 
anisotropic shapes act as sources of quenched disorder. 
On the other hand, the mechanical response is not yet well understood.
It is known that elasticity-mediated long-range interactions among
spatial inhomogeneities are crucial in systems such as 
metallic alloys \cite{Cahn,OnukiAlloy} and gels \cite{OnukiGel}.
For polydomain networks, 
the role of elastic interactions among orientationally 
correlated regions (``domains'') is yet to be clarified.
In this Rapid Communication, we numerically investigate
the mechanical response and the domain structure of model nematic networks 
incorporating both rubber elasticity and quenched random anisotropy.
Unusual soft response is obtained 
and is explained in terms of the elastic interaction.
We briefly discuss the effect of random internal stress
as another kind of quenched disorder which can destroy 
long-range orientational order \cite{Golubovic-Lubensky}.

 The total free energy of our model system is of
the form $ F = F_{el} + F_{R} + F_F $,
where $F_{el}$, $F_{R}$, and $F_F$
are, respectively, the rubber-elastic, random disorder, and
Frank contributions.
We assume networks brought deep into the nematic phase
after crosslinking in the isotropic phase, and
apply the affine-deformation theory
of nematic rubber elasticity
due to Warner et.al.\cite{WarnerOriginal}.
Then $F_{el}$ is written in terms of
the symmetric deformation tensor
$ W_{ij} = (\D r_i/\D r_k^0)(\D r_j/\D r_k^0)$,
where $ r_i^0 $ and $ r_i $
are the Cartesian coordinates of the material point
at the moment of crosslinking and after deformation,
respectively. Summation over repeated indices is implied throughout 
the paper unless otherwise stated.
It is convenient to rewrite the original form \cite{WarnerOriginal}
of $F_{el}$
using the tensor $ Q_{ij} = n_i n_j - \delta_{ij}/d $,
where $\bn$ is the director,
to obtain~\cite{Uchida-Onuki}
\eq
F_{el} = \f{\mu}{2} \int d\br ( W_{ii} 
- \alpha Q_{ij} W_{ij} ). 
\label{Fel}
\qe
The dimensionless coupling constant $\alpha (>\!0)$ is
determined by chain anisotropy and does not exceed $ d/(d-1) $.
The modulus $\mu$ is given by $k_BT$ multiplied by
the crosslink number density and
a numerical prefactor ($\sim1$) which is weakly dependent on the temperature.
We consider the incompressible limit and impose the constraint $\det\tW = 1$.
The disorder free energy 
is assumed in the form given in \cite{Clarke1,Fridrikh-Terentjev,Yu-etal},
and is rewritten as
\eq
F_{R} = \int d\br \;  P_{ij} Q_{ij},
\label{FR}
\qe
where $P_{ij}$ is a symmetric, traceless, Gaussian random tensor
with vanishing quenched average ($ \lrangle{P_{ij}(\br)} = 0 $) 
and with variance
\eq
\lrangle{ P_{ij}(\bq) P_{kl}(-\bq) } = 
U \biggl( 
\delta_{ik} \delta_{jl} + \delta_{il} \delta_{jk} 
-\f2d \delta_{ij} \delta_{kl} \biggr).
\qe
For the Frank free energy we assume the form
\eq
F_F = \f{K}{2} \int d\br \;(\grad \bn)^2.
\qe

 Here we treat the two-dimensional case for numerical 
and analytical advantages.
Then, in the absence of elasticity, our model reduces
to the random-anisotropy 2D XY model 
by regarding the unit vector $\bem=(2Q_{xx},2Q_{xy})
=(\cos 2\theta,\sin2\theta)$ as the spin variable,
where $\theta$ is the director orientation defined by
$\bn=(\cos \theta,\sin \theta)$.
We consider deformations
of the form $r_i=\lambda_i r_i^0 + u_i$ (no summation)
where $\lambda_x=\lambda$ and $\lambda_y=1/\lambda$
express the average deformation,
and $\bu=\bu(\br)$ denotes the internal displacement.
Cooling into the nematic phase tends to induce
spontaneous elongation along the director \cite{deGennesCRAS,Warner}.
If the directors are uniformly aligned along the $x$-axis,
the elastic free energy (\ref{Fel}) is minimized
at $\lambda=\lambda_m$ and $\bu=\mbox{\bf 0}$ with
\eq
\lambda_m = \biggl( \f{1+\alpha/2}{1-\alpha/2} \biggr)^{1/4}.
\qe
However, if the random field is sufficiently strong,
the system under no external stress energetically favors
a macroscopically isotropic state with $\lambda=1$.
This can be identified with the polydomain state. 
Our questions concern
how domains spontaneously deform 
and to what degree the elastic free energy 
is reduced in this highly non-uniform state.

The mechanical response was numerically simulated
by varying the macroscopic strain $\lambda$
and minimizing the free energy
with respect to $\bn$ and $\bu$ for each value of $\lambda$.
We solved the Langevin equation for the director, 
\eq
\f{\D\bn}{\D t}= ({\bm I}-\bn\bn) \cdot \lrS{ -L \,\fder{F}{\bn} + \bm{\eta}},
\qe
where $\bm{\eta}$ is an uncorrelated Gaussian thermal noise
introduced to facilitate structural evolution.
Without the noise the minimization process would stop at 
one of the local minima close to the initial configuration.
After approaching the global minimum
we turned off the noise as explained below.
The displacement $\bu$ was determined by
solving the non-linear equation
$\delta (F_{el}+ F_v)/\delta \bu =0$ with a relaxation method,
where $F_v$ is an artificial free energy functional of $\bu$,
which penalizes volume change.
In this way the local volume was kept constant 
with errors below $1\%$ throughout the runs.
Periodic boundary conditions were imposed on $\bn$ and $\bu$.
The simulation was performed on a $128^2$ square lattice with
the grid size $\Delta x=1$.
We set $K=4$ and $U=1$ for all the runs,
whereas $\mu$ and $\alpha$ were varied for different runs.
In each run the external strain $\lambda$ was slowly increased
after an initial equilibration stage at $\lambda=1$.
Occasionally, we stopped the increase of $\lambda$ and
turned off the thermal noise for an interval of time.
Thus a single run consisted of alternating periods of
annealing (with increasing strain) and quenching.
In each quench period we computed
the spatially-averaged free energy density $f=f_{el}+f_R+f_F$
and the orientation $S=\lrangle{2Q_{xx}}=\lrangle{\cos 2\theta}$.
This procedure enabled us to approximately minimize the free energy
at numerous values of $\lambda$ in reasonable computational time.
For a further check,
we then decreased $\lambda$ back from a large value
in a similar manner. A small hysteresis was obtained
but it does not affect the description below.

In Fig.1 we show the strain-stress and strain-orientation
relations for several values of $\alpha$ with $\mu \alpha^2=4$ fixed.
In both plots we can see a sharp crossover
around $\lambda=\lambda_m(\alpha)$.
Below $\lambda_m$ the total stress $\D f /\D\lambda$
is vanishingly small and slightly positive.
The average orientation increases almost linearly in the same region.
The free energy densities are plotted in Fig.2.
The elastic free energy has a slightly negative slope
in the region $\lambda<\lambda_m$,
while the disorder free energy has a positive slope and makes the
total stress slightly positive.
We chose the parameters so that
the Frank contribution is much smaller than $\mu\alpha^2$,
which is considered to be the case in typical experiments.
We also studied a few cases with stronger or weaker elastic effects.
For larger values of $\mu \alpha^2$
the shapes of the strain-elastic stress
and strain-orientation curves were almost unchanged.
For cases with $\mu\alpha^2 \simle 0.2$
these two curves exhibited less sharp crossovers.

In order to understand the origin of the soft response
it is useful to examine the structure of the polydomain
state at $\lambda=1$,
for which an analytical treatment is possible
in the weak coupling case $\alpha \ll 1$.
We expand $\Delta F_{el}=F_{el}[\bu]-F_{el}[0]$ 
with respect to $\grad \bu$ to obtain
\eq
\Delta F_{el} = \mu \int d\br 
\biggl[ \f14 \biggl( \der{u_i}{r_j} + \der{u_j}{r_i} \biggr)^2 
- \alpha Q_{ij} \der{u_i}{r_j} \biggr].
\label{DeltaFel}
\qe
Eliminating the elastic field using the conditions of
mechanical equilibrium 
$\delta \Delta F_{el}/\delta \bu = 0$
and incompressibility $\div \bu=0$,
we have a non-local elastic interaction among orientational
inhomogeneities.
We define new variables  $Q_1(\br)$ and $Q_2(\br)$ through
their Fourier transforms,
\eq
Q_1(\bq) &=& \sin (2\varphi) Q_{xx}(\bq) - \cos (2 \varphi) Q_{xy}(\bq)
,\label{Q1}\\
Q_2(\bq) &=& \cos (2\varphi) Q_{xx}(\bq) + \sin (2 \varphi) Q_{xy}(\bq) 
\label{Q1Q2},  
\qe
where $ \varphi $ is the azimuthal angle of
the wave-vector $ \bq = q(\cos \varphi, \sin \varphi)$.
Then the average free energy density reads \cite{Uchida-Onuki}
\eq
f_{el}|_{\lambda=1} = 
\mu \biggl( 1- \f{\alpha^2}{2} \lrangle{Q_1^2} \biggr)
\label{Feleff}
\qe
to order $\alpha^2$.
Note that $Q_1$ and $Q_2$ satisfy
$\lrangle{Q_1^2+Q_2^2}=\lrangle{Q_{xx}^2+Q_{xy}^2}=1/4$.
We have $\lrangle{Q_1^2}=\lrangle{Q_2^2}=1/8$
in the absence of the elastic coupling.
In its presence, 
asymmetry $\lrangle{Q_1^2}>\lrangle{Q_2^2}$ arises
to reduce the elastic free energy (\ref{Feleff}).
In the elasticity-dominated limit 
where $\mu\alpha^2$ is much larger than the disorder and
the Frank free energy densities,
we expect $\lrangle{Q_1^2} \to 1/4$,  $\lrangle{Q_2^2} \to 0$,
and $f_{el}|_{\lambda=1}\to \mu(1-\alpha^2/8)$.
Indeed these are numerically confirmed as shown in the next paragraph.
On the other hand, the elastic free energy density under
the uniform deformation with $\lambda=\lambda_m$
is also given by $\mu (1-\alpha^2/8)$ to order $\alpha^2$.
Thus, in the above limit,
the P-M transition accompanies only a small change of order $\alpha^3$
in the elastic free energy.
To see how each domain is
deformed at $\lambda=1$,
we consider the local elastic stress which is given 
as $\sigma_{ij} = \mu(\D_i u_j + \D_j u_i - \alpha Q_{ij})$
from (\ref{DeltaFel}). After some calculation, its variance 
in the mechanical equilibrium is obtained as 
\eq
\lrangle{{\sigma_{ij}}^2} = 2 \mu^2 \alpha^2 \lrangle{Q_2^2}. 
\label{varstress}
\qe
In the elasticity-dominated limit
the variance of the quantity
$\mu^{-1}\sigma_{ij}=\D_i u_j + \D_j u_i - \alpha Q_{ij}$ vanishes due to
the factor $\lrangle{Q_2^2}$ in (\ref{varstress}),
which means that each part of the system is elongated
by $1+\alpha/4 \;(\simeq\lambda_m)$ times along
the local director.
This, together with the numerical result on the mechanical response,
supports the following simple picture :
In the polydomain state
each domain is uniaxially elongated by $\lambda_m$ times
along the local director, and thus the elastic free energy is equal
to that for the monodomain state at $\lambda=\lambda_m$ (Fig.3).
The P-M transition in the region $1<\lambda<\lambda_m$
proceeds via rotation of domains
and does not change the elastic free energy.

Next we present numerical results on
the polydomain structure at $\lambda=1$,
which was studied through the correlation
function $G(r) = 2 \lrangle{Q_{ij}(\br)Q_{ij}(0)}$
and the degree of structural asymmetry $A=\lrangle{Q_1^2}-\lrangle{Q_2^2}$.
To accelerate computation of the elastic field
we assumed a weak coupling $\alpha=0.1$
and solved $\delta \Delta F_{el}/ \delta \bu=0$
under the constraint $\div \bu=0$
using FFT, instead of the relaxation method above.
The amplitude of the thermal noise was
set constant in an initial stage
and then gradually reduced to zero at a constant rate.
The correlation function is computed for the final state and
averaged over $20$ independent runs for each set of parameters.
Runs were sufficiently long
to insure that the initial configurations with
uniform and random orientations
give indistinguishable results for $G(r)$.
Shown in Figs.4 and 5 are
the correlation function and the correlation length $R$
defined by $G(R)/G(0) = 1/2$.
The elastic coupling increases the correlation length
without qualitatively affecting the form of
the correlation function.
We could not deduce a quantitative decay law for $G(r)$
from the relatively small number of samples,
but the decay was slightly faster than exponential
near the origin. For the non-elastic case the same
feature was obtained
in the Monte Carlo simulation
by Gingras and Huse \cite{Gingras-Huse}
in the presence of thermal noise,
while Yu et.al. \cite{Yu-etal} obtained exponential decay
using free boundary conditions.
Another important factor affecting $G(r)$ is the disorder strength.  
More systematic study of the decay law 
is left to future work. 
In Fig.5 the degree of asymmetry $A$ is also shown.
With increasing the magnitude of the elastic interaction 
it approaches to the upper limit $1/4$ as expected.

Finally we discuss the effect of
random internal stress 
arising from microscopic heterogeneities in the
network structure, which are intrinsic to gels \cite{Boue}.
We restrict our discussion to the case $\lambda=1$ with
small internal deformations.
In the expansion 
of the elastic free energy with respect to $\grad\bu$
there will arise an additional term,
\eq
\Delta F_{el,R} = \int d\br \; R_{ij} \;\der{u_i}{r_j},
\qe
where $R_{ij}$ is the Gaussian random stress
with $\lrangle{R_{ij}(\br)}=0$ and 
\eq
\lrangle{ R_{ij}(\bq) R_{kl}(-\bq) } = 
V_1  \delta_{ij} \delta_{kl} + 
V_2 (\delta_{ik} \delta_{jl} + \delta_{il} \delta_{jk}).
\qe
Eliminating the elastic field from $\Delta F_{el} + \Delta F_{el,R}$
we have a new interaction free energy $ \alpha\int d\br R^S_1 Q_1$,
where $R^S_1$ is defined using
the shear component $R^S_{ij}= R_{ij}- R_{kk} \delta_{ij}/d$
by an equation parallel to (\ref{Q1}) as
\eq
R^S_1(\bq) &=& 
\sin (2\varphi) R^S_{xx}(\bq) - \cos (2 \varphi) R^S_{xy}(\bq).
\qe
Treating this interaction as a weak perturbation as in~\cite{Imry-Ma},
we can see that it renders the equilibrium correlation length finite
even in the absence of the disorder free energy (3).
We mention that Golubovi\'c and Lubensky \cite{Golubovic-Lubensky}
discussed another mechanism of long-range-orientational-order 
breaking due to random stress. 
Their argument is based on the observation 
that the amplitude of thermal fluctuations 
around a uniformly aligned state diverges.
Its relevance to the present case of 
nematic networks is limited
in that their free energy does not explicitly include 
the orientational degree of freedom.

To summarize, we have numerically obtained a
soft mechanical reponse during the P-M transition. 
It originates from
structural self-organization of domains
due to the long-range elastic interaction,
and should be distinguished from
the soft elasticity \cite{Warner,Olmsted}
of uniformly oriented networks. 
The elastic contribution to the stress
is slightly negative in the transition region.
We have found a positive disorder contribution to the stress.
The elastic interaction is found to increase
the correlation length.
We have demonstrated that 
random internal stress acts as a random field on the director.
Further experimental and theoretical studies
are necessary to examine its relevance to real 
polydomain textures. 

The author gratefully acknowledges
Prof. A. Onuki for helpful discussions
and a critical reading of the manuscript,
and Dr. E. M. Terentjev for valuable comments on 
our related work.

\end{multicols} 

\begin{figure}
\epsfxsize=8cm \epsffile{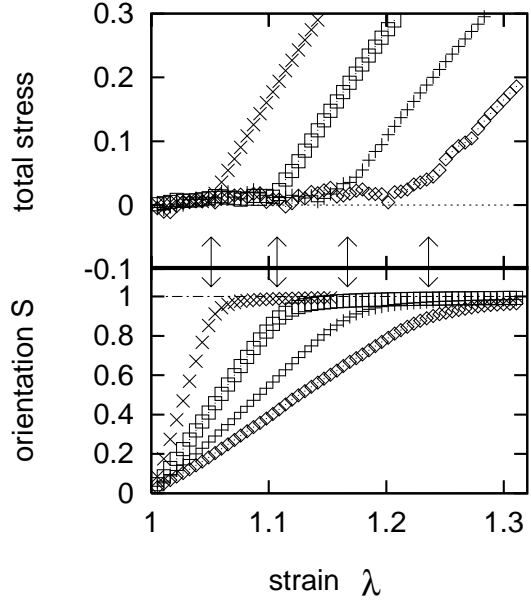}
\caption{
Top: dimensionless total stress $\mu^{-1} \D f/\D \lambda$.
Bottom: orientation $S=\lrangle{2Q_{xx}}=\lrangle{\cos2\theta}$.
Cases with different coupling strengths
$\alpha=0.2,0.4,0.6,0.8$ from left to right
are compared with $\mu\alpha^2=4$ fixed.
The arrows indicate the corresponding values of $\lambda_m$. 
}
\label{fig1}
\end{figure}

\begin{figure}
\epsfxsize=8cm \epsffile{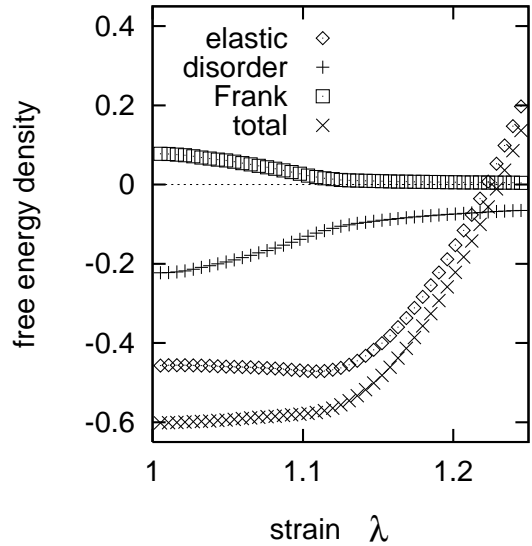}
\caption{
Free energy densities for $\mu\alpha^2=4$ and $\alpha=0.4$.
The total free energy in the polydomain regime
has a positive but small slope
due to the disorder contribution.
The value of $\mu$ is subtracted from the elastic free energy density.
}
\label{fig2}
\end{figure}

\begin{figure}
\epsfxsize=8cm \epsffile{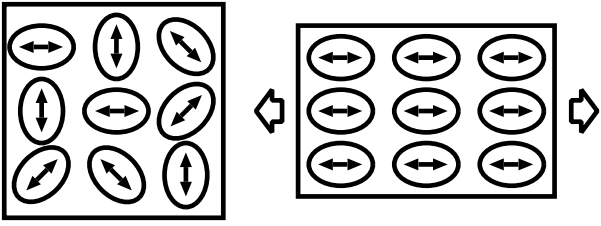}
\caption{
Schematic illustration of the P-M transition.
The ellipses represent domains under
spontaneous deformations (from circles at the moment of
crosslinking) and the arrows in them indicate the local director orientations.
Transition from polydomain at $\lambda=1$ (left) to monodomain 
at $\lambda=\lambda_m$ (right)
does not change the elastic free energy if every domain
is elongated by $\lambda_m$ times along the local director.
}
\label{fig3}
\end{figure}

\begin{figure}
\epsfxsize=8cm \epsffile{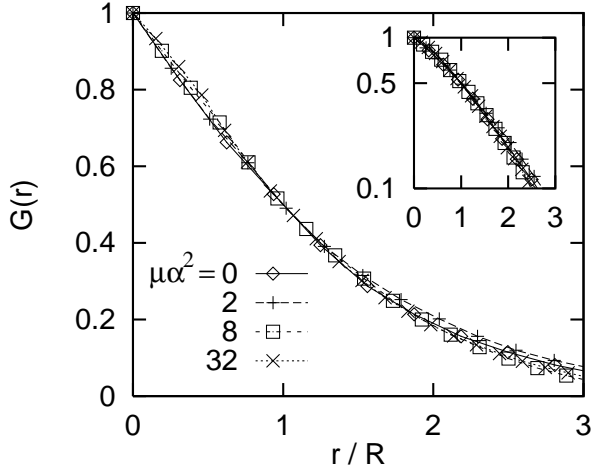}
\caption{
Correlation function $G(r)$ as a function of the
scaled distance $r/R$. 
It is insensitive to the elastic interaction.
Inset: semilogarithmic plot.
}
\label{fig4}
\end{figure}

\begin{figure}
\epsfxsize=8cm \epsffile{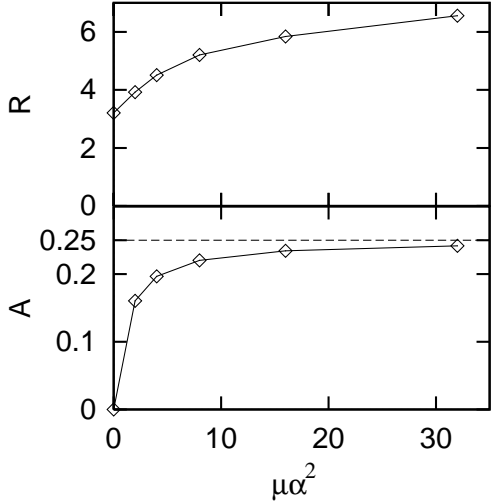}
\caption{
Top: correlation length $R$.
Bottom: structural asymmetry $A=\lrangle{Q_1^2}-\lrangle{Q_2^2}$.
}
\label{fig5}
\end{figure}

\end{document}